# Accurate distance control between a probe and a surface using a microcantilever

R. Molenaar[1], J.C. Prangsma[1], K.O. van der Werf[1], M.L. Bennink[1], C. Blum[1], V. Subramaniam[1,2]

[1] Nanobiophysics Group, MESA+ Institute for Nanotechnology and MIRA Institute for Biomedical Technology and Technical Medicine, University of Twente, PO Box 217, 7500AE Enschede, The Netherlands.

[2] Nanoscale Biophysics, FOM Institute AMOLF, Science Park 104, 1098XG Amsterdam, The Netherlands

*Keywords: nano-positioning, closed-loop feedback, drift compensation, microcantilever, AFM, hysteresis, temperature response*

We demonstrate a method to accurately control the distance between a custom probe and a sample on a µm to nm scale. The method relies on the closed-loop feedback on the angular deflection of an in-contact AFM microcantilever. High performance in stability and accuracy is achieved in this method by taking advantage of the small mechanical feedback path between surface and probe. We describe how internal error sources that find their origin in the microcantilever and feedback can be minimized to achieve an accurate and precise control up to 3 nm. In particular, we investigated how hysteresis effects in the feedback caused by friction forces between tip and substrate, can be minimized. By applying a short calibration procedure, distance control from contact to several micrometers probe-sample distance can be obtained with an absolute nanometer-scale accuracy. The method presented is compatible with any probe that can be fixed on a microcantilever chip and can be easily built into existing AFM systems.

## INTRODUCTION

Nanometer scale structures and effects based on the nanometer proximity between objects play an increasingly important role in scientific research and applied nanotechnology. This is the driving force behind the development of techniques to control the position of objects with nanometer accuracy. Many techniques to achieve this goal have been developed, relying on, for example, capacitive or interferometric sensors to perform feedback [1,2]. The wide spectrum of situations where accurate positioning has to be achieved, spanning from research in life sciences to semiconductor industrial processes, comes with many different preconditions and technical challenges. These require a large diversity of methods for accurate position control, each with their own specific advantages.

In this paper we present an instrument that controls the distance between a probe and a surface with nanometer accuracy over a micrometer range. Though our instrument can be used in many situations where the distance between two objects has to be controlled on such scales, the motivation for this work originates from a nanophotonic application. In this application the fluorescence lifetime of emitters is modified by positioning a mirror in close proximity (ranging from in contact to a distance of ~ 1000 nm) to the fluorophores[3,4]. From the measured relation between fluorescence lifetime and mirror-fluorophore distance, important photophysical properties such as the radiative and non-radiative decay rate and the fluorescence quantum efficiency can be obtained [5, 6, 7, 8, 9].

To obtain sufficient accuracy there are a few important requirements that the control of the mirror-sample distance has to fulfill. Firstly the accuracy in positioning should approach < 5 nm over a distance range of 1000 nm. Secondly since low signals will require measurement times of 10 to 1000 seconds a good long-term stability needs to be obtained. Additionally, because the physical properties of the fluorophores we want to relate with the mirror distance are derived from optical measurements, there are several extra requirements and constraints. Most importantly, the distance control has to be implemented on top of an existing (commercial) microscope employing a water or oil immersion objective. This also implies the use of typically 0.17 mm thick coverslips as sample substrates. We stress that these requirements and constraints are very common in biological applications where high-resolution imaging and accurate mechanical manipulation with a custom probe often are combined. Our specific application requires the probe to be a highly reflective surface, which we realized by using a 100 μm diameter spherical mirror. However, the operation of the device allows the use of other suitable probes, e.g. non-reflective functionalized or sharp probes.

Typically optical microscopes are not designed for mechanical stability in the nanometer range. Temperature stability is for instance low because the design is typically not balanced for time constants of the thermal expansion. Also the sample itself, consisting of a standard thin glass coverslip, has a low mechanical stability and is dynamically deformed by capillary forces of the water or oil immersion objective due to evaporation and focusing. At these scales thermal drift, vibrations and van der Waals surface-probe interactions become non-negligible factors that need to be compensated. Clearly it is not straightforward to control the probe-sample distance in this system at the nanoscale level without a proper feedback mechanism or physical contact of the probe itself.

Our design is based on a feedback mechanism on the deflection of an in-contact microcantilever in an atomic force microscope. AFM [10, 11, 12] is a well-known tool for mapping sample topography by measuring heights with a lateral resolution on the nanoscale. Although the objective of both tapping mode and contact mode AFM is to measure nanoscale details of the substrate, the tilt between the sample and scanner, as well as any mechanical drift of the total system during scanning also contributes to the measured height signal. Importantly, to remove the unavoidable mechanical drift in the height that occurs on the time scale of the formation of a single image, AFM images are real-time filtered by a high-pass filter or offline filtered by image-processing software. This means that though an AFM is excellent in maintaining distances as long as tip and sample interact, it is in its classical form not capable of maintaining or controlling larger distances when out of contact. However, in contact the AFM can sense and control nanoscale distances making it a perfect platform to act as a real-time mechanical feedback control actuator. To do this successfully the AFM design needs to be adapted to make it less sensitive to low frequency drift, enable correction for mechanical drift, and to operate beyond the typical deflection range.

Literature reports several drift compensation methods using the AFM: for precise nano-manipulation, a straightforward method is to track the surface position at a reference position and repeat this over time [13]. This method however does not assure a real-time accuracy in the probe-sample distance unless the entire setup has been optimized for extreme stability. Altman et al [14] introduced a dual microcantilever system, in which the long secondary microcantilever assists the approach of the primary shorter cantilever, when the latter is not in-contact with the substrate surface. Because the primary microcantilever still can bend freely, this method cannot compensate for long-range electrostatic forces and van der Waals attraction forces. Alternatively, feedback methods based on shear forces make use of the frequency detuning of the oscillating tuning fork that senses shear forces with the surface, limiting its range to distances less than 10 nm [15].

We developed a method in which the deflection of the AFM microcantilever is used to control the distance between the fixed probe and the substrate with nanoscale accuracy, as shown in figure 1. The distance to be controlled is the distance between mirror probe and sample indicated as d in the figure. To realize this approach, the mirror's lowest point is aligned in the horizontal plane with the microcantilever tip by fixing the mirror at distance $S = H_{mirror}/\sin(\alpha_{tip})$ from the tip. Importantly, the mirror is attached rigidly to the stiff microcantilever chip and has a high mechanical stability. By tilt-adjustment with the AFM-head tripod, the mirror is brought in proximity to the surface, while the microcantilever is already in-contact and having a, for AFM standards, large dynamic range in deflection of 2 μm. The microcantilever deflection is now a direct measure of the mirror-surface distance with nanoscale accuracy. In the closed loop feedback method we are able to achieve distance control with: (1) minimized drift over time, (2) a linearized movement after calibration, (3) displacement range from in-contact up to 2μm, and (4) a positioning accuracy of better than 3nm.

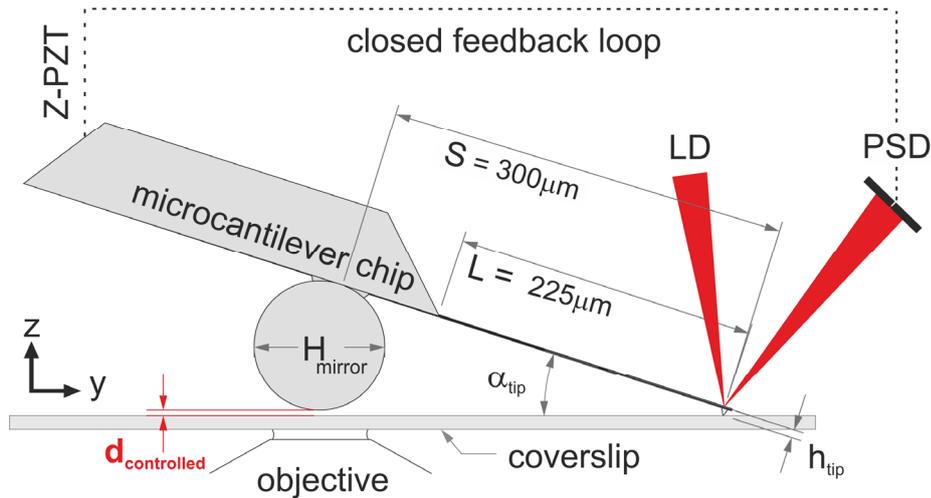

FIG. 1. Schematic of the feedback concept. Microcantilever with a spherical mirror rigidly attached to the chip is brought into firm contact. The microcantilever angular deflection as measured by the reflected spot of the laser diode (LD) on the position sensitive detector (PSD) is used as sensor to control the distance between the mirror and surface indicated by $d_{setpoint}$. The angle between the surface and the microcantilever chip is $\alpha_{tip}$. L is the microcantilever length. S is the distance between the mirror center and the microcantilever tip. $H_{mirror}$ is the mirror diameter. $h_{tip}$ is the tip height.

This manuscript is organized as follows: section II identifies typical drift sources on the measurement platform and positioning errors that arise in the microcantilever use. Section III describes the implementation and experimental performance of the nano-positioning with the AFM system and performance of drift compensation. Finally, section IV presents conclusion and recommendations.

**II OVERVIEW OF ERRORS THAT AFFECT THE DISTANCE**

The error sources in the distance between surface and mirror can be separated into two classes. Firstly there are external effects that can be compensated and motivate our use of a feedback system. These effects are mainly induced by mechanical instabilities in the construction and the largest amplitude drifts are associated with long term (>1s) drifts. Typical magnitudes of these drifts are discussed in subsection A. Secondly, there are internal effects, since the overall accuracy of any feedback cannot be more accurate than the internal reference. Drift occurring in the AFM internal angular detection system consisting of a laser diode [16], position sensitive diode (PSD) and buffer electronics are not compensated and lead to drift in the separation distance. We identified 3 error sources that are an intrinsic part of our feedback mechanism and that, although they cannot be compensated, can be minimized in our design. Mechanical vibrations and the effect of non-flatness of the surface are discussed in subsection B. The effect of the bimetallic temperature response of the microcantilever on positioning is discussed in subsection C. Positioning precision in the feedback is optimized by reducing the microcantilever hysteresis and microcantilever buckling originating from surface-tip friction as explained in subsection D.

**A. Experimental mechanical limitations**

The microcantilever with attached mirror is positioned in the AFM-head on top of a confocal microscope. Drift arises in this system because both the AFM-head and the microscope are constructed of parts with different thermal expansion coefficients, thermal capacities, and mechanical tension. To characterize the typical drift between the AFM-head and the coverslip on the confocal microscope, we brought the microcantilever in-contact with the coverslip. Typical mechanical drift is shown in figure 2 (a) (black line), where the system drift after the first warmup hour is on the order of 80nm/h. The initial ramp-up is caused by a stretching of the Z-Piezoelectric transducer (PZT-Z) due to heating by the internal laser diode, this possibly causes some erratic quick changes. When the water immersion objective is brought in-contact with the coverslip (red line), the capillary interaction of the evaporating water film makes drift over time up to 5 times larger than that observed for the air objective. Most importantly, since coverslip displacement is induced by the capillary forces, refocusing of the objective onto the coverslip has dramatic effects on the distance. Figure 2 (b) shows that displacements on the order of 100 nm when refocussing are not uncommon. The above discussed sources of displacement can be compensated by the feedback system we propose, resulting in two orders of magnitude improvement in positioning precision.

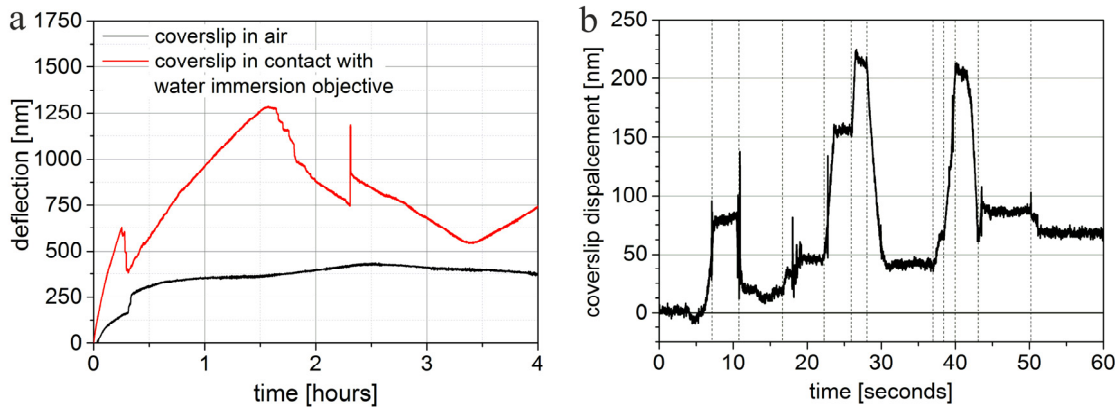

FIG. 2. (a) Separation measured by the AFM deflection positioned on the confocal microscope in-contact with the coverslip. A typical Z-direction drift occurring over a 4 hour time period for a 170 μm thin coverslip positioned in air (black line, color online) and in-contact with the water immersion objective (red line, color online). Deflection sensitivity is calibrated from voltage to nanometers by driving the PZT-Z with a known displacement. (b) Z displacement of the coverslip induced by re-focusing by a water immersion objective, yielding on average a 100 nm displacement. The dotted vertical lines mark the moment when the change of focusing starts.

**B. Mechanical vibrations and substrate topography**

Two essential mechanical limitations of the surface on which the experiment is performed are of influence on the accuracy obtained: acoustic vibrations and the topography of the substrate. First, from our experience we observe that the 170 μm coverslips are quite sensitive to vibrations induced by acoustic sources. We measured that acoustic vibrations can lead to up to 10 nm peak-to-peak amplitudes of the coverslip. Thus for a precise Z positioning, acoustic noise sources need to be minimized. Second, any tilt of the coverslip results in small vertical displacement from lateral movement of the stage, which is fully compensated by the feedback. However, the sample surface, in our case a clean coverslip, is not flat. Due to the ~300 μm lateral separation between the microcantilever tip and the mirror, lateral variations in height of the sample can lead to variations in absolute height and thereby loss of precision when the sample or mirror are moved laterally. To estimate the magnitude of this effect in our application, we measured the flatness of the coverslip on a 1/20 λ flat reference glass. We find a typical slope of 500 nm/mm with a derivative in the slope in the order of 150 nm·mm$^{-2}$. Thus the slope variance in our system that has a 300 μm separation between tip and mirror is 45 nm. This means that to stay within the 3 nm accuracy limit we are limited to lateral movements less than 66 μm away from the point where the calibration is performed.

**C. Microcantilever temperature response**

The microcantilever, whose angular deflection acts as the distance sensor, is the main mechanical construction between mirror and coverslip. In many commercially available microcantilevers the back side is standard coated with an aluminum or gold film to enhance the optical beam reflection. This reflective coating results in a bi-metal behavior of the microcantilever, making the feedback system dependent on the temperature of the environment [17, 18]. Though the feedback keeps the deflection signal constant it cannot discriminate between changes in deflection induced by temperature or displacement. For this reason we measured and optimized the temperature response of the used microcantilever by recording the deflection change when it is approached from a far distance > 10 mm towards an aluminum housed electronic resistor that is kept at temperature 15 °C warmer than room temperature. For a Bruker MSCT tip-D with a 45 nm Au coated backside, the temperature induced deflection is measured as 80 nm/K towards the surface, that falls within the range found in literature of 50 nm/K [19] to 166 nm/K [20]. To compensate for this bimetallic effect we sputtered a 60 nm Au layer on the front side of the microcantilever chip, which lead to a minor reduction of the temperature response to 40 nm/K. As a next step, to reduce it furthermore we used commercial uncoated microcantilever tips (Bruker, MSCT-UC) which we then coated on both sides with a thin 4 nm Cr adhesion layer and 80 nm Au in a balanced in-house sputter machine. In this way a minimized temperature response of 5 nm/K is achieved reducing the temperature sensitivity 16-fold with respect to a standard MSCT tip-D. With this design a 3 nm accuracy corresponds to a temperature fluctuation in the laboratory of ΔT <0,6 °C within the timeframe of the measurement, which can easily be achieved. Note that an alternative method for reducing the temperature dependence would be to use a stiffer microcantilever. However, the discussion in the next section about hysteresis problems will clarify why this is not a suitable solution.

**D. Minimization of microcantilever hysteresis**

It is important to realize that, as a result of the 1-2 μm Z displacement, the tip slides a few hundred nanometers on the surface along the Y direction (see figure 1 for YZ-axis). A simple geometric argument shows that this is well approximated by $\Delta y = -\Delta Z \cdot \tan(\alpha_{tip})$ [21]. The resulting sliding movement and the friction between surface and tip causes microcantilever buckling, due to the moment resulting from the lateral fiction force, illustrated in figure 3A. This leads to a hysteresis in the deflection that can be observed when the tip is moved up and down through one cycle. The hysteresis acts within the feedback system as a highly undesired phenomenon resulting in overshoot of the PZT-Z and mirror in the feedback with the PZT-Z compensation.

The microcantilever buckling we discuss as an adverse effect is actually used to measure local surface friction coefficients in lateral force microscopy. Palacio *et al* [22], reviews several techniques to calibrate deflection signals in lateral force microscopy. From this work, the following two things can be learned: firstly, a larger microcantilever tip height creates a larger moment enhancing the buckling effect. Second, the normal spring constant $k_Z$ and lateral spring constant $k_Y$ originate from the microcantilever shape, geometry and material properties. The microcantilever lateral spring constant $k_Y$ is never specified by the manufacturer but can be calculated [23, 24, 25, 26] and is about $2 \cdot 10^2$ times larger than $k_Z$ for a beam shaped microcantilever vs. $1.5 \cdot 10^3$ times larger for a triangle shaped microcantilever (see figure 4 for microcantilever shapes).

The order of magnitude difference in $k_y$ between triangular and beam shaped cantilevers means triangular microcantilevers are much less prone to buckling, leading to a drastic reduction of the hysteresis. Also the spring constant $k_z$ is of importance. With a low spring constant, the normal force $F_N$ the tip exerts on the substrate is minimal, thus minimizing the friction force $F_{FR}$ which is dependent on the normal force $F_N$ between tip and sample. A reduction in friction directly leads to a reduction of hysteresis in the feedback. However, because the total normal force $F_N$ that a microcantilever tip exerts on the surface is the sum of the capillary forces $F_{CAP}$ and cantilever forces $F_C$, the normal force $F_N$ has a practical minimum given by the attractive capillary forces. These capillary forces are in the order of tens of nN, thus choosing a microcantilever with a spring constant $k_z$ smaller than 0.01 N/m does not lead to significant reduction of the hysteresis.

To demonstrate that the hysteresis results from differences in the normal load, we measured the hysteresis of a microcantilever array (Bruker, MSCT) and a beam shaped microcantilever (Nanoworld, FMR). These experiments are performed on a mica surface on a metal disk. In Figure 3B, we show the force-distance curve for the MSCT tip-D and FMR microcantilever. To overcome the attractive capillary forces to release the MSCTtip-D, the AFM needs to pull 380 nm from the surface. The capillary forces at displacement $Z = 0$ therefore pulls the tip on the surface with $F_{CAP} \sim 8nN$. The maximal cantilever force at Z displacement +700 nm is $F_{C-max} = 21$ nN. With a low ratio $F_{C-max}/F_{CAP}$ the cantilever lateral friction and thus hysteresis remains minimal as a function of microcantilever deflection.

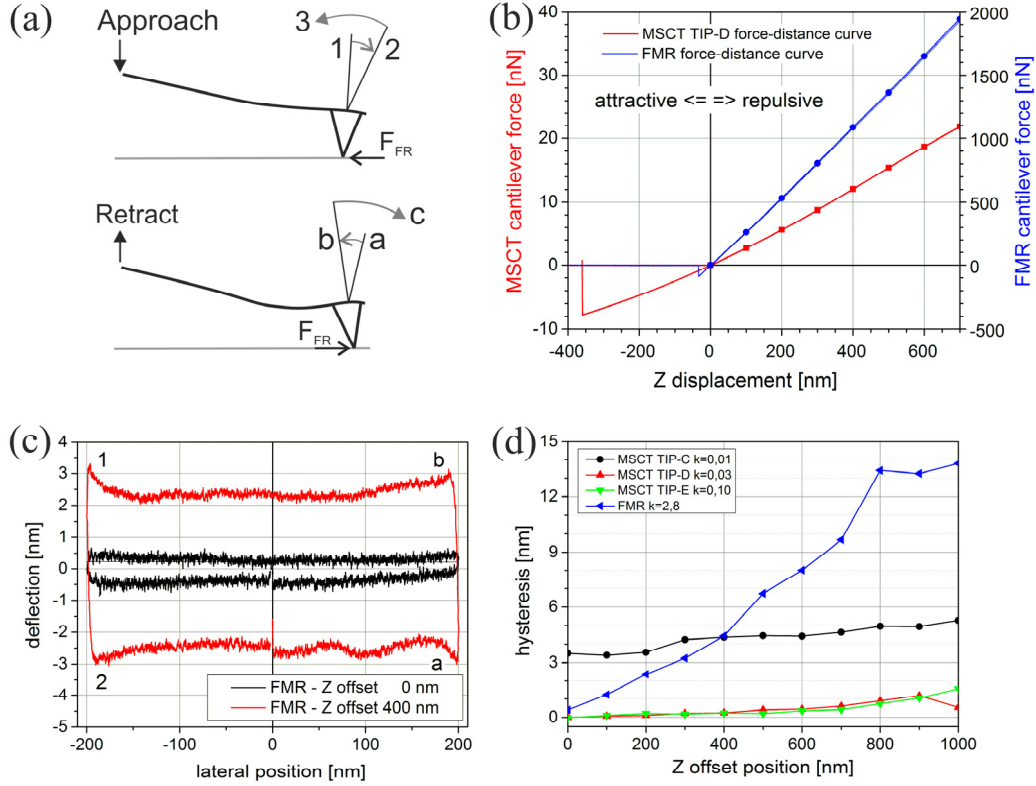

FIG. 3. (a) Schematic depiction of the hysteresis source, the friction moment between the tip and surface buckles the cantilever tip deflection in opposite direction after change in Z positioning direction, indicated by the arrow between 1-2 and a-b. The turnover is when this hysteresis zone has overcome, deflection proceeds as normal depicted by arrow 2-3 and b-c. (b) Typical AFM force distance curves for a small spring constant cantilever (MSCT tip-D, left axis, red) and a large spring constant (FMR, right axis, blue). For negative Z displacement the attractive adhesion force is visible. The markers display the Z offset positions where the lateral scan is performed to determine the hysteresis results shown in figure d. (c) Hysteresis measurement of FMR tip by scanning the AFM PZT in lateral Y direction over 400 nm back and forward. Average over 16 cycles is shown that were recorded at 2 Hz. Red line (color online) Z offset at 400 nm (large normal force), black line Z offset at 0 nm (minimal normal force). (d). Experimental result of increasing deflection hysteresis as function of the Z offset from 0 to 1000 nm, comparing 4 microcantilevers with different spring constants and geometries.

The hysteresis is systematically studied by lateral scanning the microcantilever 400 nm to and fro in Y direction and plotting the angular deflection value as a function of its lateral position. Hysteresis is defined as the difference between the average deflection from the to and fro scan. For each microcantilever the in-contact position (Z = 0 nm) is determined and from this point the sample stage positioner (Physik Instrumente, PI-527.3CD) is brought closer with 100 nm incremental steps to Z = 1000 nm. So the hysteresis is measured as a function of increasing repulsive force. In figure 3C, we show as example the FMR tip at Z step 0 nm and step 400 nm where we find respectively 1 and 5 nm hysteresis in deflection. In Figure 3D, the results of this analysis for the 4 studied microcantilever tips are shown. For the low spring constant microcantilever MSCT tip-C with k = 0.01 N/m we find a constant hysteresis of 4 nm over the full Z range. We explain this by the attractive adhesion force $F_{CAP}$, which is on the order of 20-50 nN and therefore dominant over the repulsive cantilever force which is maximally 10 nN at 1000 nm deflection.

For the MSCT tip-D with k = 0.03 N/m we observe a slight increase in hysteresis from 0 to 2 nm as expected where the repulsive force is maximally 30 nN. For the FMR tip k = 2.8 N/m the repulsive cantilever force rises to more than 2800 nN. In this case capillary forces are not significant anymore. The microcantilever deflection for this tip shows irregular deflection from Z = 500 nm on linear motion that cannot be reproduced, possibly due to the large normal forces. For the FMR hysteresis is rapidly increasing from 0.5 nm at 0 nm to beyond 14 nm at 1000 nm bending. This high spring constant microcantilever is thus not suitable as feedback sensor due to increasing hysteresis.

From the above we conclude that a low axial spring constant $k_z$ with a strong lateral spring constant $k_y$ is necessary for minimal hysteresis. Because normally only the spring constant $k_z$ is specified, a strong lateral spring constant $k_y$ can be maximized by selecting a microcantilever with a large width, low tip height and triangular shape. Simple lateral force measurements can be used to characterize hysteresis and aid the selection of the most optimal microcantilever for the feedback system.

## III    REALIZATION OF THE INSTRUMENT

The best concept for positioning the mirror with high accuracy is based on the results obtained from the previous section. The mirror, in this our case a 100 μm diameter polystyrene sphere with minimal surface roughness [27] was fixed on the AFM microcantilever chip with UV-curing glue applied with a micropipet on a hydraulic micromanipulator. The stiff connection prevents that the mirror is moved towards the surface by attraction of the surface's van der Waals forces and long range electrostatic forces. By fixing the mirror to the microcantilever chip, the mirror and height feedback sensor are brought as close together as possible, eliminating most of the relevant drift sources in the mechanical construction. The cantilever remains in contact, providing the feedback signal at all times. Prior to the experiment we link the microcantilever deflection angle to the distance separation in a calibration procedure. For realization of the concept we adapted a custom-built AFM [28].

### A. Closed loop feedback

As feedback sensor for the closed loop concept we chose the 225 μm length microcantilever, (Bruker MSCT-UC tip-D) because the deflection will offer us the required dynamic range of 2 μm. Also, the positioning precision of this microcantilever is optimal due to the minimized hysteresis in the feedback. The microcantilever chip is mounted in the AFM-head with a tilt of 18°. From a simple geometrical consideration: $S = (H_{mirror} - h_{tip})/\sin(\alpha_{tip})$ the 100 μm diameter Polystyrene (PS). The mirror was positioned at ~ 300 μm distance from the microcantilever tip. The total mechanical path length from the mirror to the coverslip is thus reduced to less than 400 μm. This reduction to a very short path length is an essential aspect of the drift reduction in our design. When the microcantilever is in contact with the coverslip, the feedback loop is closed, and the microcantilever deflection value can be held constant by the AFM feedback system, achieving real-time drift compensation during the experiments.

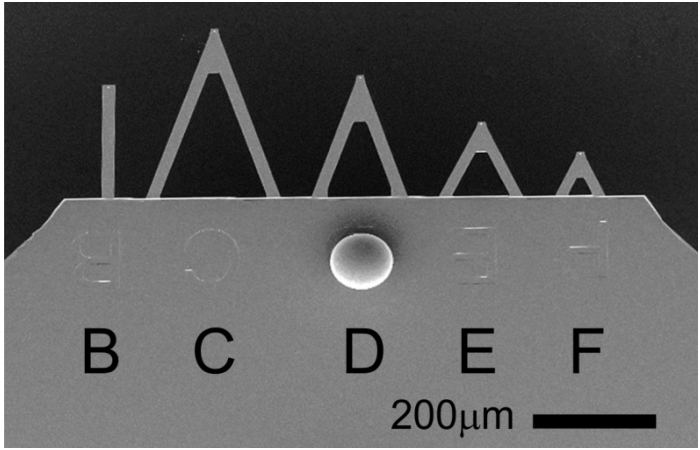

FIG. 4. SEM image of a microcantilever chip, with a 102μm PS sphere fixed in line with a Bruker MSCT tip-D and sputtered with a 60 nm Au layer.

## B. Calibration procedure

A typical AFM is actually designed to operate in shorter deflection range instead of our required 2000 nm. In our case the angular deflection is used in a regime where a part of the light falls off the PSD and the sum value drops. Normalization of deflection voltage with the sum voltage doubles the linear range in which the setup can be used and makes the deflection sensitivity independent on laser power fluctuations. The resulting normalized deflection-distance curve is however not completely linear with Z distance but exhibits a S-shape. Therefore, a calibration procedure is necessary to convert the normalized deflection scale of the microcantilever to a relative nanometer scale. The most straightforward method would be to use an AFM PZT-Z equipped with a capacitive feedback system. Then the calibration procedure could be performed independently with only the AFM-head. However, in our setup the AFM-head is equipped with an open-loop PZT-Z. For the required calibration we therefore used the calibrated and capacitive feedback controlled sample scanner (Physik instrumente, P-527.3CD) that holds our sample. In the calibration procedure we link the microcantilever deflection to the Z motion of the calibrated sample scanner. In a first step we obtain a relative distance calibration, with an at first unknown offset. To determine the absolute position between mirror and surface we need to detect when the mirror has reached the surface. To determine when the mirror touches the sample surface we use the discrete change in the deflection sensitivity of the microcantilever that appears during the calibration procedure from the moment that the mirror contacts the surface, see figure 5. We further observed that at this point the confocal laser spot at the glass interface becomes defocussed, which indicates the possible deformation of the coverslip explaining why there is a continuous change in cantilever deflection after contact. The direct contact movement is executed with precise control (<30nm) by use of the sample scanner and we found this deformation to be fully reversible and non-intrusive for our application. If absolute measurements are of importance for more delicate probes other reference/stop criteria such as a discrete electronic current upon contact can be implemented.

In practice, we perform the calibration by linear displacement of the sample stage with attached coverslip. During movement the capacitive Z sensor position and microcantilever deflection are recorded, resulting in the calibration curve shown in figure 5. For absolute distance control the reference is marked from the discrete slope transition. To allow fast software-controlled feedback the measured deflection is converted to nanometers by use of polynomial coefficients, which are extracted from the calibration curve by a polynomial curve fit.

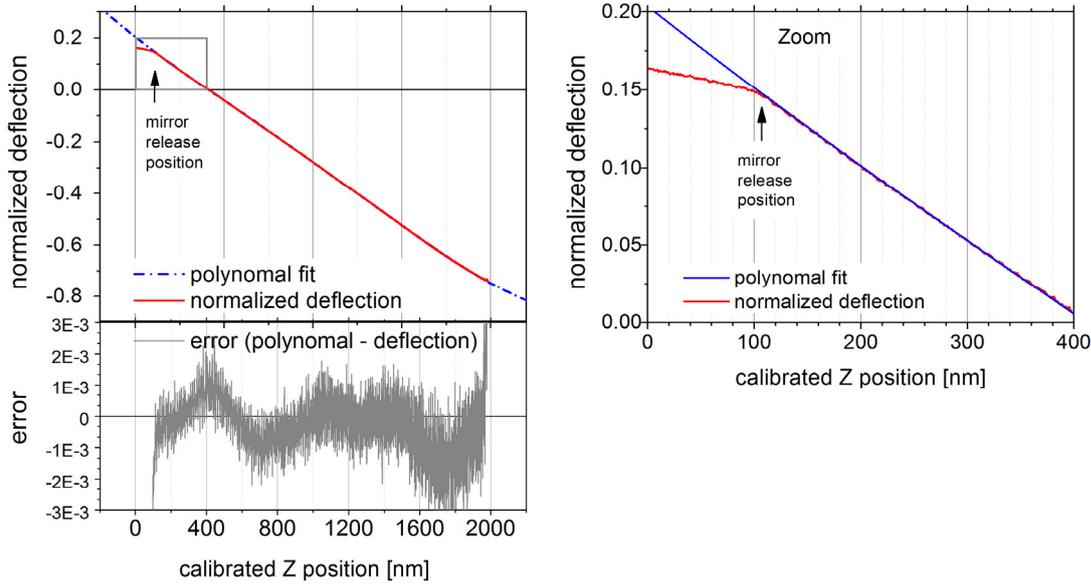

FIG. 5. (a) typical calibration curve, normalized deflection vs. relative position. The cantilever is in contact at position 0 nm and, the calibrated stage moves away from the cantilever generating a non-linear deflection. Error between the deflection signal and its polynomial fit, that remains within 2 nm error. (b) Zoom-in on the calibration curve, showing the sudden change in slope when the mirror leaves the sample surface.

## C. Digital feedback implementation

The nature of our experiments requires the feedback system, to control mirror-sample distance accurately up to 1000 seconds. We chose a feedback scheme that mainly compensates the long-term externally induced drift, and for this the feedback bandwidth is set to 30Hz (video rate) to enable compensation by user input and rapid repositioning. Due to the low frequency bandwidth it is possible to perform the feedback on the software level. The remaining high frequency noise is about 3nm peak-to-peak and is attributed to the remaining hysteresis in the feedback and the acoustic noise. For many applications these small variations will average out. Onboard electronics in the AFM convert the signal from the PSD to a deflection and sum voltage. Both signals are registered by a data acquisition card (National Instruments, PCIe-6353) that samples the deflection and controls the AFM Z-PZT voltage. Control by Labview software simplifies the implementation of the calibration curve and dynamic distance control in the experiment. The software timed feedback loop runs at a rate of 2.5 kHz. In each interval the timed loop samples the deflection, determines the error compared to the set point, and a PID algorithm compensates the error by setting the compensation on the analog output that is brought back through a high voltage amplifier to the AFM PZT-Z actuator.

### D. Demonstration of positioning system

After the calibration procedure is executed, the feedback is immediately engaged to keep the distance to a preset distance. Any waveform, within the feedback bandwidth of 30Hz, can be applied as input to the set point. To verify the positioning resolution we apply a 3 nm step size discrete motion. From the initial 2 seconds, system noise is measured as 0.62 nm rms, shown in Figure 6.

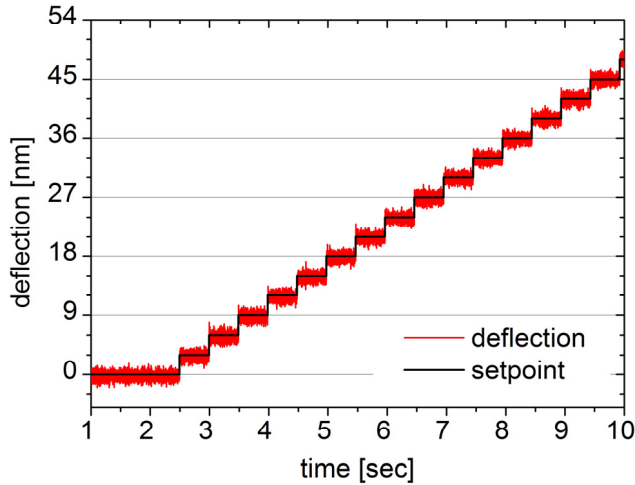

FIG. 6. Typical performance of the AFM feedback. Incremental set point steps of 3 nm are clearly followed by the calibrated deflection feedback.

### E. Drift detection by external reference

To exclude the presence of any processes or errors that lead to a change in distance undetected by our feedback system based on an in-contact AFM cantilever, we performed an independent measurement to validate the distance stability of our feedback system. To do this we image the interference rings that are formed between the mirror and coverslip that are a direct measure of the distance between sample and mirror. To quantify possible drift, we limit ourselves to the linear response regime of the intensity on displacement of an interference fringe to a ± 30 nm range. We thus limit the use of the external reference as a sensor to detect if the mirror distance remains in position when the feedback is engaged.

To imaging the mirror interference rings we modified the confocal microscope. As a lightsource of monochromatic (525nm, 5 nm bandwidth) light we used a fiber coupled white light laser (Fianium, SC400-pp) equipped with an acousto-optic tunable filter for wavelength selection. The collimated output from the fiber was focused onto the back focal plane (BFP) of the objective via the back-port of the microscope. To overlap the excitation and emission beam in the filter cube, the dichroic mirror was replaced with a beam sampler wedge (10% reflection). Interference between the glass/air interface of the coverslip and the spherical mirror gives rise to ring shaped interferences due to the increasing distance of the mirror curvature. This interference pattern was imaged on a EMCCD camera (Andor, ixon DU897-BV). A 760 nm short-pass filter cuts off the residual light form the AFM diode. Due to the illumination of sample and mirror via the BFP, the interference pattern is insensitive for objective focusing. Laser power stability is measured within 0,7% and the laser intensity was reduced with OD filters and the camera EMCCD gain is disabled. All frames were recorded with a 14 ms exposure time.

The interference pattern between the flat coverslip and curved mirror results in a gradient in interference periods. If no drift is present, the inference intensity at a single fringe will be stationary. Nanometer displacement results in a movement of the interference pattern (see figure 7).

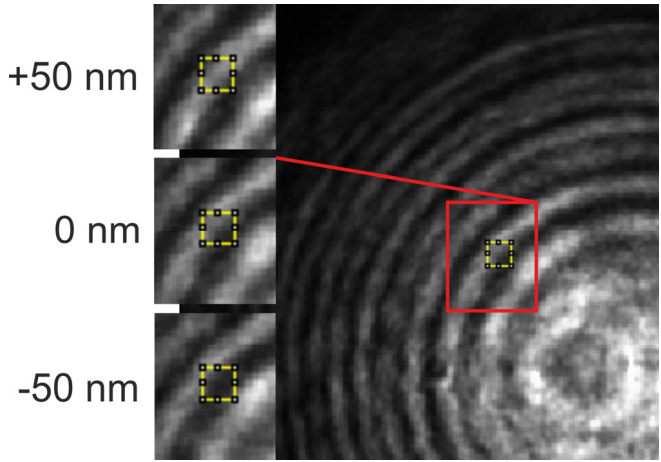

FIG. 7. A quarter section of the ring interference image by the camera, the intensity is recorded in the yellow 9x9pixel box. The 3 inserts show the fringe intensity pattern as result of the calibrated mirror Z displacement.

Before the stability test the mirror is swept by the PZT-Z over ±50 nm with 10 nm increments and the interference pattern is recorded. To avoid any drift resulting from piezo relaxation and creep from the AFM-PZT-Z, no voltage was applied on the actuator during the no-compensation measurement. Mirror displacement was measurement over 30 minutes with an interference interval of 5 seconds, with and without feedback compensation. For the analysis we selected a box of 9x9 pixels corresponding to a half of the local fringe distance to maximize intensity response and therefore the spatial accuracy.

The initial calibration displacement with 10 nm increments shows a linear intensity response in the 9x9 box. From the intensity response the corresponding displacement can be extracted. Figure 8, shows the mirror displacement measured with local fringe intensity with and without feedback compensation. We tested the feedback over 30 minutes and found a long term stability of 2.5 nm (adjacent average over 30 points) with a standard deviation of 1.5 nm towards the adjacent average. With the feedback disabled the system drifts beyond 40 nm with a standard deviation of 0.9 nm towards the adjacent average. The increased noise level for feedback compensation is explained from the shorter exposure time of 14 ms. that is faster than the set feedback bandwidth of the mirror.

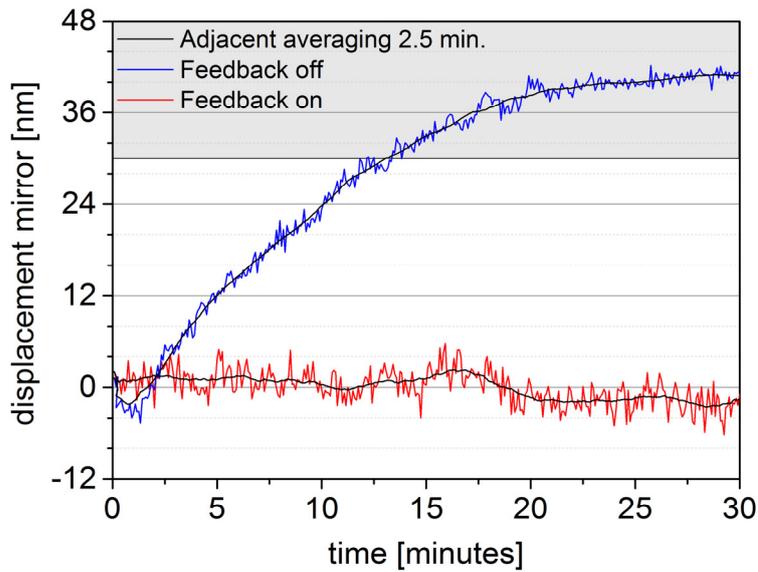

FIG. 8. Coverslip-mirror distance in nanometers, recorded by interference fringe intensity over 30 minutes. The values in the gray area fall out of the 30 nm linear range and values are not representable anymore, yet indicate drift is ongoing.

## IV CONCLUSION

We have demonstrated the use of a calibrated feedback controlled deflection on an AFM system that can be used as a nm accurate distance control of a non-contact mirror, with a positioning range of a few micrometers. This enables us to measure photophysical properties relying on an active compensation for drift on time scales > 30 ms. We showed that a temperature balanced cantilever decreases thermal response, allowing operation in normal laboratory environments. A careful analysis of the hysteresis effects occurring in the tip-sample interaction is presented giving insight in the parameters that lead to the best feedback conditions. We find that optimal performance is achieved with a low spring constant microcantilever tip to reduce surface friction, together with a microcantilever whose geometrical shape is minimally sensitive to buckling. Calibration of the deflection is performed by the sample scanner with capacitive feedback. Experimental resolution drift compensation was found to be within our 3 nm specifications and the long term distance stability was verified using interferometric measurements. The design can be flexibly and non-invasively integrated with standard microscopes making it a versatile platform for accurate distance control with many nanoscale application.

## ACKNOWLEDGEDGMENTS

The authors acknowledge support from the Dutch Technology Foundation STW, project number 12149, which is part of a STW Perspective program on Optical Nanoscopy.

**REFERENCES**

[1] Y. Ganjeh, B. Song, K. Pagadala, K. Kim, S. Sadat, W. Jeong, K. Kurabayashi, E. Meyhofer and P. Reddy, Rev Sci Instrum 83, 105101 (2012).

[2] J. White, H. Ma, J. Lang and A. Slocum, Rev Sci Instrum 74, 4869 (2003).

[3] B. C. Buchler, T. Kalkbrenner, C. Hettich and V. Sandoghdar, Phys Rev Lett 95, 063003 (2005).

[4] K. H. Drexhage, B Am Phys Soc 14, 873 (1969).

[5] Y. Cesa, C. Blum, J. M. van den Broek, A. P. Mosk, W. L. Vos and V. Subramaniam, Phys Chem Chem Phys 11, 2525 (2009).

[6] C. Blum, Y. Cesa, M. Escalante and V. Subramaniam, J R Soc Interface 6, S35 (2009).

[7] C. Blum, N. Zijlstra, A. Lagendijk, M. Wubs, A. P. Mosk, V. Subramaniam and W. L. Vos, Phys Rev Lett 109, 203601 (2012).

[8] P. Lunnemann, F. T. Rabouw, R. J. A. van Dijk-Moes, F. Pietra, D. Vanmaekelbergh and A. F. Koenderink, Acs Nano 7, 5984 (2013).

[9] A. I. Chizhik, I. Gregor, B. Ernst and J. Enderlein, Chemphyschem 14, 505 (2013).

[10] G. Binnig, C. F. Quate and C. Gerber, Phys Rev Lett 56, 930 (1986).

[11] H. J. Butt, B. Cappella and M. Kappl, Surf Sci Rep 59, 1 (2005).

[12] R. Garcia and R. Perez, Surf Sci Rep 47, 197 (2002).

[13] C. Spagnoli, A. Beyder, S. R. Besch and F. Sachs, Rev Sci Instrum 78, 036111 (2007).

[14] S. M. Altmann, P. F. Lenne and J. K. H. Horber, Rev Sci Instrum 72, 142 (2001).

[15] T. R. Albrecht, P. Grutter, D. Horne and D. Rugar, J Appl Phys 69, 668 (1991).

[16] R. Kassies, K. O. van der Werf, M. L. Bennink and C. Otto, Rev Sci Instrum 75, 689 (2004).

[17] S. Singamaneni, M. C. LeMieux, H. P. Lang, C. Gerber, Y. Lam, S. Zauscher, P. G. Datskos, N. V. Lavrik, H. Jiang, R. R. Naik, T. J. Bunning and V. V. Tsukruk, Adv Mater 20, 653 (2008).

[18] L. Wu, T. Cheng and Q. C. Zhang, Measurement 45, 1801 (2012).

[19] L. A. Wenzler, G. L. Moyes and T. P. Beebe, Rev Sci Instrum 67, 4191 (1996).

[20] J. R. Barnes, R. J. Stephenson, C. N. Woodburn, S. J. Oshea, M. E. Welland, T. Rayment, J. K. Gimzewski and C. Gerber, Rev Sci Instrum 65, 3793 (1994).

[21] J. L. Hutter, Langmuir 21, 2630 (2005).

[22] M. L. B. Palacio and B. Bhushan, Crit Rev Solid State 35, 261 (2010).

[23]J. L. Choi and D. T. Gethin, Nanotechnology 20**,** 065702 (2009).

[24]M. Muller, T. Schimmel, P. Haussler, H. Fettig, O. Muller and A. Albers, *Surf Interface Anal,* (2006).

[25]Y. L. Wang and X. Z. Zhao, Rev Sci Instrum 80**,** 023704 (2009).

[26]R. J. Warmack, X. Y. Zheng, T. Thundat and D. P. Allison, Rev Sci Instrum 65**,** 394 (1994).

[27]P. J. van Zwol, G. Palasantzas, M. van de Schootbrugge, J. T. M. de Hosson and V. S. J. Craig, Langmuir 24**,** 7528 (2008).

[28]K. O. Vanderwerf, C. A. J. Putman, B. G. Degrooth, F. B. Segerink, E. H. Schipper, N. F. Vanhulst and J. Greve, Rev Sci Instrum 64**,** 2892 (1993).